\documentclass[runningheads]{llncs}
\usepackage{graphicx}
\usepackage{ebproof}
\usepackage{amssymb}
\usepackage{latexsym}
\usepackage{graphicx}
\usepackage{comment}
\begin{document}
\title{Automated ZFC Theorem Proving with E}
%
%
\author{John Hester}
\authorrunning{J. Hester }
%
\institute{Department of Mathematics, University of Florida, Gainesville, Florida, USA
\url{https://people.clas.ufl.edu/hesterj/}\\
\email{hesterj@ufl.edu}}
\maketitle              
\begin{abstract}
I introduce an approach for automated reasoning in first order set theories that are not finitely axiomatizable, such as $ZFC$, and describe its implementation alongside the automated theorem proving software E.  I then compare the results of proof search in the class based set theory $NBG$ with those of $ZFC$.

\end{abstract}
\section{Introduction}

Historically, automated reasoning in first order set theories has faced a fundamental problem in the axiomatizations.  Some theories such as $ZFC$ widely considered as candidates for the foundations of mathematics are not finitely axiomatizable.  Axiom schemas such as the schema of comprehension and the schema of replacement in $ZFC$ are infinite, and so cannot be entirely incorporated in to the prover at the beginning of a proof search.  Indeed, there is no finite axiomatization of $ZFC$ \cite{zfcnotfinite}.

As a result, when reasoning about sufficiently strong set theories that could no longer be considered naive, some have taken the alternative approach of using extensions of $ZFC$ that admit objects such as proper classes as first order objects, but this is not without its problems for the individual interested in proving set-theoretic propositions.

As an alternative, I have programmed an extension to the automated theorem prover $E$ \cite{E} that generates instances of parameter free replacement and comprehension from well formuled formulas of $ZFC$ that are passed to it, when eligible, and adds them to the proof state while the prover is running.  This allows directly reasoning in $ZFC$, avoiding the problems of reasoning in other theories.  By using a fair algorithm for selecting replacement and comprehension instances, every possible such instance will eventually be generated given infinite time and resources.  This means that refutational completeness will be preserved as long as every possible comprehension and replacement instance is eventually fed to the prover.    

\section{$ZFC$ and $ZFC^o$}

$ZFC$ and $NBG$ are theories built on first order logic.  $NBG$ is a conservative extension of $ZFC$ when only discussing sets \cite{nbgcons}.  Most presentations of $NBG$ allow for special quantifiers that quantify only over sets, but the first order formulation uses the set predicate.  As a helpful reference, I describe the relevant theories here.  In the experiments section, I have used TPTP encodings of these axioms, the $NBG$ axioms provided by Art Quaife \cite{quaifebook}\cite{quaifenbg}.  Below are the axiom schemas of $ZFC$ and $ZFC^o$, I have omitted the rest of the axioms as they can be found in nearly any book on set theory.

\subsection{Axiom Schemas of $ZFC$}

\subsubsection{Axiom Schema of Comprehension:}  Let $\phi (x,v_1,...,v_n)$ be any formula in the language of $ZFC$ with free variables among $x,v_1,...,v_n$, and let $y$ be some some variable not in $\phi$.  Then

    $$\forall v_1 ... \forall v_n\forall a\exists y\forall x (x\in y \leftrightarrow x\in a \wedge \phi(x,v_1,...,v_n))$$

\subsubsection{Axiom Schema of Replacement:}  For every formula $\phi(x,y,v_1,...,v_n)$ of the language of $ZFC$,

$$\forall v_1...\forall v_n (\forall x\exists y\forall y'(\phi(x,y',v_1,...,v_n)\leftrightarrow y'=y) \rightarrow$$ $$\forall a \exists b\forall y (y\in b \leftrightarrow\exists x \in a\  \phi(x,y,v_1,...,v_n)).$$

\subsection{$ZFC^o$}

$ZFC^o$, or parameter free $ZFC$, is an alternative axiomatization of $ZFC$ where the schemas of comprehension and replacement have been replaced by their parameter free counterparts, and the rest of the axioms remain the same.  $ZFC^o$ is equivalent to $ZFC$ as every instance of the full axioms of comprehension and replacement can be derived in a finite number of steps in $ZFC^o$ \cite{zfc0}.  

\subsubsection{Parameter Free Schema of Comprehension:}  Let $\phi(x)$ be any formula in the language of $ZFC$ with a single free variable $x$, and let $y$ be some variable not in $\phi$.  Then

$$\forall a\exists y \forall x (x\in y \leftrightarrow x\in a\wedge\phi(x))$$

\subsubsection{Parameter Free Schema of Replacement:}  For every formula $\phi (x,y)$ of the language of $ZFC$,

$$\forall x\exists y \forall y' (\phi(x,y')\leftrightarrow y'=y)\rightarrow$$ $$\forall a\exists b\forall y(y\in b \leftrightarrow\exists x\in a\ \phi(x,y)).$$

\section{Downsides of $NBG$ for Automated Foundations}
Art Quaife uses the theory $NBG$ (for Neumann-Bernays-G{\"o}del) in his book "Automated Development of Fundamental Mathematical Theories" \cite{quaifebook}.  While $NBG$ admits a finite axiomatization, and provides a richer ontology as well, it has other properties that may hamper proof search.  

\subsection{Sethood as a Predicate}  For instance, in $NBG$ the first order objects are all classes, while sets are just a special type of class.  This means that every quantifier is quantifying over a universe that the mathematician may not be interested in, and whose objects may not have sufficient structure to say much.  In other words, there is too much expressive power.  

It is true that $NBG$ is a conservative extension of $ZFC$ when the domain of discourse is restricted to just sets, but doing so has the negative side effect of imposing new predicates on every formula in consideration that demand everything being discussed is a set and not a proper class.  Such predicates may be equivalently expressed as $x\in V$ where $V$ is the class of all classes (which is an object of $NBG$) or simply as $Set(x)$.  

As a result, a simple set-theoretic proposition such as $\exists x\phi (x)$ that simply wishes to assert the existence of a set satisfying a certain formula becomes $\exists x (x\in V \wedge \phi(x))$.  The presence of these extra predicates, the corresponding increase in the scope of quantifiers, and increased size of expressions complicates the work of automated theorem provers.  In addition, the nearly universal presence of this sethood predicate could make the given clause function's job of selecting relevant clauses more difficult.

\subsection{Approximation of Comprehension}

In the absence of the axiom schema of comprehension, $NBG$ set theory has a class existence theorem asserting that the collection of objects satisfying a formula with parameters forms a class.  However, this theorem is technically a theorem schema as there is the same problem of asserting the existence of an object for every $\phi$ as is the case with the schema of replacement and comprehension in $ZFC$.  So, when considering $NBG$ as an alternative to $ZFC$, in some sense the problem of not having a finite axiomatization has an analog in the theory. 

The structure of this state of affairs is not necessarily conducive to automated theorem proving.  Statements asserting the existence of collections satisfying certain properties are indispensable in mathematics and it is difficult to expect the automated prover to reproduce the steps taken in the proof of the class existence theorem for particular formulas while trying to prove a conjecture that needs it.  Because of the theorem status of class existence for general formulas, the job of the automated theorem prover is much more difficult, unless particular instances of class existence are included as axioms.

\section{Implementation of Axiom Schemas as Inference Rules}

The schemas of parameter free replacement and parameter free comprehension can be interpreted as the below inference rules, where wff is an abbreviation for well formed formula.

    \[
  \begin{prooftree}
    \Hypo{\phi(x) \mbox{ is a wff}}
    \Hypo{y \mbox{ does not occur in }\phi(x)}
    \Infer2{\forall a\exists y \forall x (x\in y \leftrightarrow x\in a\wedge\phi(x))}
  \end{prooftree}
    \]
    
        \[
  \begin{prooftree}
    \Hypo{\phi(x,y) \mbox{ is a wff}}
    \Infer1{$$\forall x\exists y \forall y' (\phi(x,y')\leftrightarrow y'=y)\rightarrow$$ $$\forall a\exists b\forall y(y\in b \leftrightarrow\exists x\in a\ \phi(x,y)).$$}
  \end{prooftree}
    \]

\subsection{Fragmentary Approach}There were two approaches taken in the project.  The first and simpler approach is to generate the parameter free comprehension and parameter free replacement instances corresponding to every eligible clause generated in the proof search, then add them to the proof state.  In both cases, the axiom schemas of parameter free replacement and parameter free comprehension are replaced by inference rules that take an input clause and return the corresponding replacement or comprehension instance if possible.   This is easy to check, as if there is one free variable, you know there is a corresponding comprehension instance, and if there are two variables you know there are corresponding replacement inferences. 

This is done by adding the clauses generated by the schema inference rules to the \verb|tmp_store| of the proof state, which imitates the process by which new clauses are added to the collection of unprocessed clauses during a normal E proof search.  While this produces $ZFC$ proofs and benefits from the internal guidance in E by applying inference rules to the desirable clauses selected by the given clause algorithm, it has a serious downside as this will only produce a fragment of $ZFC$.  Only applying the inference rule to the clauses generated in proof search will mean that there are many clauses and formulas that are never generated and so will never have their corresponding replacement and comprehension instances added to the proof state.

\subsection{Full Approach}
The second and more thorough approach is to generate and maintain a list of many well formed formulas of $ZFC$, and generate more after all of them have been used for replacement or comprehension instances.  This is possible because the well formed formulas of $ZFC$ are recursively enumerable.  As the proof search is ongoing, the prover can evaluate a fraction of the list of well formed formulas for relevance, and choose the formula with the best score.  This can be done during every loop of the given clause algorithm, or much less frequently.  Once this is done, the selected formula has the parameter free replacement and comprehension inference rules applied to it and any generated clauses are added to the proof state as before.

As we can theoretically generate every well formed formula of $ZFC$ given enough time and resources, and every instance of the schema of parameter free replacement and parameter free comprehension is a consequence of one of the above inference rules, we can gain equivalence to $ZFC$ as long as every one of these well formed formulas are eventually selected using a fair approach.  An example of such an approach would be to alternate selecting well formed formulas for generating comprehension and replacement instances based off of relevance and based off of the order they were generated in.  However, in practice it is useful to sacrifice this equivalence by focusing on relevance because the formulas selected by the order they were generated may be useless for the current proof search and create unnecessary unprocessed clauses.  

This approach is essentially automated theorem proving in a very large theory, so the techniques used in that field of research could be very fruitful.  In practice, generating a large number of well formed formulas and deciding which ones are useful for creating schema instances is not an easy problem, so in the experiments section below I focused on applying schema generating rules to the clauses generated during proof search that will hopefully be most relevant to the problem at hand.

\section{Experiments}

The \verb|SET| directory of the TPTP library contains a large number of set theory problems, many of them in the language of $NBG$ \cite{TPTP}.  In order to compare the approaches and merits of $NBG$ and $ZFC^o$, I have taken $124$ of the $NBG$ problems and corresponding definitions, and transformed them in to the language of $ZFC$.  This mostly entails removing predicates from the $NBG$ statements that assert certain objects are sets, as this is unnecessary in the language of $ZFC$, so the corresponding $ZFC$ problems are simpler to express.  From a theoretical point of view, since $NBG$ is a conservative extension of $ZFC$, and $ZFC^o$ is equivalent to $ZFC$, for the chosen problems every proof that is found in $NBG$ should have a corresponding proof in $ZFC^o$.  In all of the proof attempts described here I took the fragmentary approach described in the previous section, so full equivalence is lost.

Often, it turns out that the proofs in $ZFC$ are shorter than corresponding proofs in $NBG$, sometimes much shorter.  This seems to be due to the fact the axiomatization of $ZFC$ removes the need to verify that some objects of interest are sets.  Because the schema instances fit a common pattern, it is common for $E$ to introduce many new definitions that are only used in a schema instance and so were effectively useless.  Towards reducing the number of definitions, the best $ZFC$ performance on many $TPTP$ problems was obtained using the $E$ options \texttt{--no-eq-unfolding --definitional-cnf=100}.  Below is a graph comparing the proof lengths of $TPTP$ problems using the \texttt{--auto} mode of $E$ in $ZFC$ and $NBG$ both with and without the definitional options, for the problems which at least one version of $ZFC$ and $NBG$ could find solutions.

\hspace*{-2cm}\includegraphics[]{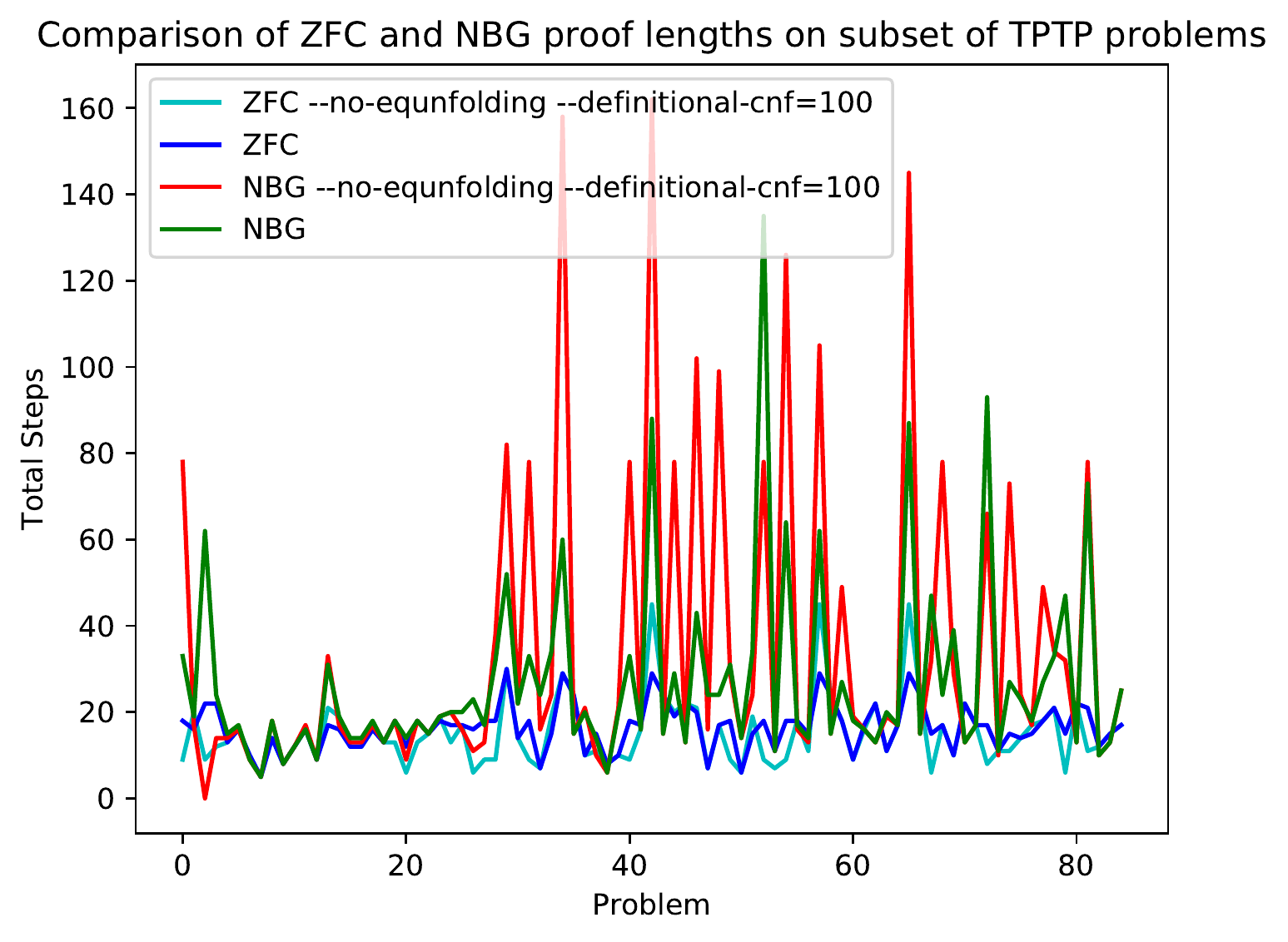}

In addition, I created a custom problem set of $30$ more interesting set theoretical propositions, dealing with ordinals, cardinalities, and bijections.  On this collection, the difference in proof length was negligible compared to the TPTP set problems, but the $ZFC$ systems had a higher percentage of problems solved.  In this setting, the definitional options in fact hampered the proof search.  In total, with both testing sets brought together, the percentage of problems solved by $ZFC$ and $NBG$ were very similar despite changes in definitional behavior.  $ND$ below denotes the theory with the definitional options mentions above.

\begin{table}[]
\centering
\begin{tabular}{lllll}
\hline
                                         & ZFC  & ZFC-ND & NBG  & NBG-ND \\ \hline
Percentage of selected TPTP solved       & 67\% & 77\%   & 77\% & 72\%   \\
Percentage of custom set problems solved\footnotemark & 60\% & 50\%   & 40\% & 40\%   \\
Total percentage                         & 47\% & 70\%   & 72\% & 67\%   \\ \hline
\end{tabular}
\end{table}

\footnotemark[1]{Best ZFC performance on the custom set theory problmes was obtained in a slightly different version of the project in which schema instances were printed then fed to the E parser, rather than built with E's internal data types.}

\section{Future Work}

In the approach taken above, the axiomatization chosen was the parameter free $ZFC^o$.  This was done for practical reasons.  The full comprehension and replacement schemas remain metatheorems in $ZFC^o$, in parallel to the class existence metatheorem of $NBG$.  Further experiments with full comprehension and replacement as inference rules may provide more performance, as in the current approach one must still derive many comprehension and replacment instances that would be necessary for interesting conjectures.

As there are many possibilities for comprehension and replacement instances to be added to the state, this problem can be compared to the issue of theorem proving with very large axiom lists.  It would be interesting to use the approaches taken in research on automated theorem proving on large theories, such as ENIGMA \cite{ENIGMA} or using a watchlist approach such as ProofWatch \cite{ProofWatch}.  In addition, machine learning approaches such as those found in Deep Network Guided Proof Search \cite{DNG} could provide increased performance by selecting only the axiom instances that are necessary.  The value of this cannot be understated as the approaches presented here add many unprocessed clauses to the state that are not necessarily helpful.

The Mizar project uses Tarski-Groethendieck set theory as its foundation, which is itself an extension of $ZFC$.  Formal proofs available through Mizar could provide an invaluable source of training data for an automated theorem prover implementing Tarski-Groethendieck set theory in a way very similar to what is described in this paper.

\section{Conclusion}

I have programmed an extension to the E prover that allows automatic generation of the schemas of parameter free replacement and parameter free comprehension as inference rules rather than axioms.  In the presence of the finite number of other axioms, this allows automated theorem proving to be done for the first time in a fragment of $ZFC$ containing arbitrarily many instances of the schemas of comprehension and replacement.  I also have provided a description of the theory $NBG$ and compared it with parameter free $ZFC$ as a foundational theory for automated theorem proving.  Using the axioms of $ZFC$ allowed the automated proof of 15 difficulty 1.0 problems from $TPTP$'s \texttt{SET} directory.

Almost all theorems proved by $ZFC^o$ in the experiment could also be proved by $ZFC$ with the axiom schemas removed, and in some cases the many new unprocessed clauses prevented the prover from finding a proof while using the theory $ZFC^o$.  This indicates that many problems problems in the \texttt{SET} directory of the $TPTP$ library either did not need the axiom schemas, or were too difficult to solve even with them.  This is also true of the custom set theory problems I used in the experiments section.  Custom problems that were intentionally formulated to need the axioms of schema or replacement also failed.  It is likely that the same problem of the class existence theorem being a metatheorem in $NBG$ is paralleled in $ZFC^o$ by the necessity of derivations for the full comprehension and replacement instance to be found.

$ZFC$ theorem proving on general problems seemed to be very comparable to that of $NBG$ in success rates, but also provided much shorter proofs in some situations.  This suggests that with improved guidance functions and the full schemas of comprehension and replacement, $ZFC$ based automated proof attempts could yield more successes than $NBG$.  In particular, $ZFC$ seemed to have better performance on deeper problems that dealt with more complex predicates. 

\section{Acknowledgements}  I would like to thank Douglas Cenzer, Josef Urban, and Martin Schulz for helpful conversations that led to this experiment.
%
%
%
%
%
%

\end{document}